\documentclass[12pt]{article}
\usepackage{tikz-cd}
\usepackage{tikz-cd}
\usepackage{authblk} 
\usepackage{graphicx}%
\usepackage{multirow}%
\usepackage{amsmath,amssymb,amsfonts}%
\usepackage{amsthm}%
\usepackage{mathrsfs}%
\usepackage{pgfplots}
\usepackage{tikz}
\usepackage{longtable}
\usetikzlibrary{arrows.meta, positioning}
\usepackage[title]{appendix}%
\usepackage{xcolor}%
\usepackage{textcomp}%
\usepackage[colorlinks=true, linkcolor=blue]{hyperref}
\usepackage{tabularx}
\usepackage[linesnumbered,lined,boxed,commentsnumbered]{algorithm2e}
\usepackage[noend]{algpseudocode}
\usepackage{mdframed}
\usepackage{tikz}%
\usepackage[
n, 
advantage,
operators,
sets,
adversary,
landau,
probability,
notions,
logic,
ff,
mm,
primitives,
events,
complexity,
oracles,
asymptotics,
keys
]{cryptocode} 
\usetikzlibrary{positioning, arrows.meta}%
\newtheorem{theorem}{Theorem}
\newtheorem{definition}{Definition}

\theoremstyle{thmstyleone}%
\theoremstyle{thmstyletwo}%
\theoremstyle{thmstylethree}%
\def\Z{{\mathbb Z}}
\def\Q{{\mathbb Q}}
\def\N{{\mathbb N}}
\def\F{{\mathbb F}}
\def\C{{\mathbb C}}

\def\a{{\mathfrak a}}

\def\m{{\mathsf m}} 

\def\g{{\mathfrak g}}
\def\id{{\mathsf{id}}}
\def\param{{\mathsf{param}}}
\def \C{\mathcal{C\ell(O)}}
\def \O{\mathcal{O}}
\def \Ell{\mathcal{E\ell\ell}_p{(\mathcal O)}}
\def\key{\bgroup\key@}
\usepackage{graphicx}

 \newcounter{case}

 \renewcommand{\thecase}{\arabic{case}}

\newcounter{subcase}

 \renewcommand{\thesubcase}{\alph{subcase}}

\providecommand{\keywords}[1]{\textbf{\textit{Keywords:}} #1}
\usepackage{color}

\begin{document}
\title{CSI-IBBS: Identity-Based Blind Signature using CSIDH}
%
%

\author[1]{Soumya Bhoumik\thanks{Corresponding author: \href{mailto:soumyabhoumik@example.com}{s\_bhoumik@fhsu.edu}}}
\author[1]{Sarbari Mitra}
\author[2]{Rohit Raj Sharma}
\author[2]{Kuldeep Namdeo}

\affil[1]{Department of Mathematics, Fort Hays State University, Hays, KS, USA}
\affil[2]{Department of Mathematics, Maulana Azad National Institute of Technology, Bhopal, India}






\maketitle              

\begin{abstract}
Identity-based cryptography (IBC), proposed by Adi Shamir, revolutionized public key authentication by eliminating the need for certificates, enabling a more efficient and scalable approach to cryptographic systems. Meanwhile, in \cite{Katsumata2024group}, Katsumata et al. were the first to present the blind signature protocol based on the hardness assumption of isogeny with provable security, which resembles the Schnorr blind signature. Building upon these foundational concepts, we propose an Identity-Based Blind Signature Scheme with an Honest Zero-Knowledge Verifier utilizing the CSIDH framework. This scheme combines blind signatures for privacy preservation with zero-knowledge proofs to ensure the verifier's honesty without revealing any additional information.

Leveraging the quantum-resistant properties of CSIDH, a post-quantum secure scheme based on supersingular isogenies, our scheme offers strong protection against quantum adversaries while maintaining computational efficiency. We analyze the security of the introduced protocol in the standard cryptographic model and demonstrate its effectiveness in safeguarding privacy and verifier honesty. Furthermore, we present a performance evaluation, confirming the practical viability of this quantum-resistant cryptographic solution for privacy-preserving applications. This work advances the creation of secure, and scalable cryptographic systems for the post-quantum era.
\end{abstract}


\section{Introduction} 

The rise of digital communication and e-commerce has driven the need for strong cryptographic designs to ensure security and authenticity. Blind signature schemes are crucial in this context, enabling anonymous and untraceable transactions in applications like digital cash, e-voting, and private authentication. These schemes permit a signer to sign a message without any information about its content, preserving user privacy. When combined with zero-knowledge proofs, they further enhance protocol integrity by ensuring verifier honesty. Identity-based cryptography ($\mathsf{IBC}$), introduced by Shamir in 1985 \cite{shamir1985identity}, simplifies public key management by using identities as public keys, eliminating the need for certificates. This makes identity-based signatures ($\mathsf{IBS}$) efficient for resource-constrained environments like wireless sensors. Classical $\mathsf{IBS}$ schemes—such as Fiat-Shamir \cite{fiat1986prove} and Guillou-Quisquater \cite{girault1991identity}, leveraging on the hardness of the discrete logarithm problem, is now become susceptible to quantum attacks. As quantum computing advances, research has been shifted towards quantum-resistant $\mathsf{IBS}$ protocol based on lattice assumptions \cite{tian2013efficient, tian2014efficient}.\\
In the evolving landscape of post-quantum cryptography (PQC), isogeny-based cryptography has grabbed significant attention for its potential to withstand quantum attacks. A pioneering effort in this area was the identity-based signature protocol introduced by Galbraith et al. \cite{galbraith2017identification}, which leveraged the Supersingular Isogeny Diffie-Hellman (SIDH) framework. This line of research gained further momentum with the introduction of the Commutative Supersingular Isogeny Diffie-Hellman (CSIDH) protocol by Castryck et al. in 2018 \cite{castryck2018csidh}, sparking interest in adapting classical cryptographic schemes to the CSIDH setting.

Several notable signature schemes have since been developed based on CSIDH, including SeaSign \cite{de2019seasign} and CsiFish \cite{beullens2019csi}, both of which employ the Fiat-Shamir transform \cite{fiat1986prove}. For instance, CSIDH-512 offers 127 bits of classical security and 64 bits of quantum security \cite{castryck2018csidh}. These constructions benefit from compactness, though they remain computationally intensive due to the ideal class group action, which involves computing chains of isogenies.

A significant advancement in this domain is CsiIBS \cite{peng2020csiibs}, an identity-based signature scheme that eliminates the need for public key certificates by using identities directly as public keys. This approach streamlines authentication in post-quantum environments and employs a variant of the Fiat-Shamir transform tailored to isogeny-based identification schemes. The result is a compact signature format with strong security guarantees, including existential unforgeability against chosen identity and message attacks ($\mathsf{UF\text{-}CMA}$). Addressing certain limitations in earlier schemes, Shaw and Dutta \cite{shaw2021identification} refined key components such as the sampling and Extract algorithms, while also providing more rigorous and consistent security proofs to enhance both efficiency and robustness.

In a parallel direction, Katsumata et al. \cite{Katsumata2024group} introduced the first provably secure isogeny-based blind and partially blind signature scheme. Their construction leverages the quadratic twist of elliptic curves and a novel ring variant of the GAIP. This scheme achieves compact signature sizes and introduces a fresh approach to partial blindness that avoids traditional hashing techniques.

This paper presents the design, implementation, and analysis of an Identity-Based Blind Signature ($\mathsf{IBBS}$) scheme that merges the blind signature framework of CSI-Otter \cite{Katsumata2024group}, which employs honest verifier zero-knowledge (HVZK) proofs, with the identity-based identification protocol by Shaw and Dutta \cite{shaw2021identification}. Our scheme leverages the identity-based structure for simplified key management and authentication, while preserving the blindness and HVZK properties of CSI-Otter to ensure user privacy and secure verification. Built on the CSIDH framework, the proposed $\mathsf{IBBS}$ scheme is quantum-resistant, computationally efficient, and well-suited for scalable, real-world applications that demand both privacy and strong security guarantees. We detail the construction of the scheme, analyze its security under the standard model, and demonstrate its practicality through performance evaluations, highlighting its potential in shaping a resilient, privacy-preserving, and quantum-secure cryptographic future.

\subsection{Outline of this paper}
The structure of this article is as follows. In section \ref{sectiomPrelim}, we have elaborated the required mathematical background and cryptographic foundations, including an overview of isogeny-based cryptography, identity-based signatures, and blind signature schemes. Section \ref{sectionIBID} is about the Identity-Based Identification ($\mathsf{IBID}$) scheme, which adapts Shaw and Dutta's identity-based signature scheme \cite{shaw2021identification} within the framework of vectorization, quadratic twists, and CSI-SharK. Section \ref{sectionIBBS} refers to our proposed design of the Identity-Based Blind Signature ($\mathsf{IBBS}$) protocol, detailing its components and operational workflow. Section~\ref{sectionSecurity} offers a comprehensive security analysis, demonstrating the resistance of both schemes to classical and quantum attacks, and includes performance evaluations to illustrate their practicality. Section~\ref{performance} discusses the computational costs and signature sizes associated with our schemes. This paper concludes in Section~\ref{conclusion}, where we summarize our contribution and suggest directions for future work.

\section{Preliminary}\label{sectiomPrelim}

\subsection{Notations and Abbreviation}

We outline here the notation and conventions employed throughout the paper. $\N$ and $\Z$ have been used in their usual form to denote the sets of natural numbers and integers, respectively. A summary of the remaining notations used throughout this article are given in \autoref{tab:Notions_table} and abbreviations used are given in \autoref{tab: abbreviations_table}
\renewcommand{\arraystretch}{1.3}
\begin{longtable}{|>{\raggedright\arraybackslash}p{3.5cm}|>{\raggedright\arraybackslash}p{10cm}|}
\caption{Key Mathematical Notations Employed in this Work} 
\label{tab:Notions_table}\\
\hline
\textbf{Notion} & \textbf{Description} \\
\hline
$\Z_N$ & ring of integers modulo $N$ with elements chosen from the range $[-N/2, N/2) \cap \Z$. \\
\hline
$[k]$ & set $\{1, 2, \dots, k\}$ for $k\in \Z^+$ \\
\hline
$\mathbf{v}[i]$ represents the  & Vector containing its first $i$ elements \\
\hline
$v_i$ & i-th entry of vector $\mathbf{v}$\\
\hline
 $x \xleftarrow{\$} S$. & Uniform sampling of a random element $x$ from $S$\\
\hline
$\odot$  & Element-wise multiplication of vectors in $\mathbb{R}$ \\
\hline
$\parallel$   & String concatenation \\
\hline
$g^{\mathbf{a}}$  & $(g^{a_1}, \dots, g^{a_n})$\\
\hline
$\ast$ & Mathematical operation\\
\hline
$g^{\mathbf{a}} \ast h^{\mathbf{b}}$ & $(g^{a_1} \ast h^{b_1}, \dots, g^{a_n} \ast h^{b_n})$\\
\hline
$\Sigma$-protocol & Sigma protocol\\
\hline
$\C$ & Ideal class group\\
\hline
\end{longtable}
\vspace{10pt}

\renewcommand{\arraystretch}{1.3}
\begin{longtable}{|>{\raggedright\arraybackslash}p{3.5cm}|>{\raggedright\arraybackslash}p{10cm}|}
\caption{Table of Abbreviations}
\label{tab: abbreviations_table}\\

\hline
\textbf{Abbreviation} & \textbf{Description} \\
\hline
PQC & Post-Quantum Cryptography \\
\hline
IBC & Identity-based Cryptography\\
\hline
IBBS & Identity-based Blind Signature \\
\hline
CSIDH & Commutative Supersingular Isogeny Diffie–Hellman \\
\hline

IBID & Identity-based Identification \\
\hline
HVZK &  Honest Verifier Zero-Knowledge \\
\hline

\end{longtable}

\subsection{Elliptic Curves and Isogenies}
An elliptic curve is a smooth, projective algebraic curve of genus $1$ defined by a cubic equation of the form $y^2= x^3+ax+b$ ( where $a,b \in \mathbb{F}_p $ ) and the condition $4a^3+27b^2 \neq 0$. The set of points satisfying the above equation forms an abelian group with respect to point addition, and the point at infinity serves as the identity.

 Consider the elliptic curve  $E_1(\mathbb{F}_p)$ and $E_2(\mathbb{F}_p)$ having $p>5$ so that the $char (\F_p) \neq 2,3$ and let $\mathcal{O}_{E_1}$ and $\mathcal{O}_{E_2}$ are points at infinity of the respective elliptic curves $E_1$ and $E_2$. The isogeny is a non-constant surjective morphism  $\phi : E_1 \longrightarrow E_2$,
 satisfying the condition $\phi(\mathcal{O}_{E_1})= \mathcal{O}_{E_2}$.  
 An isogeny's degree refers to the degree of the associated rational map between elliptic curves. For separable isogenies, the degree is equal to the cardinality of the kernel.  $\#ker(\phi)$ is always finite, so $ker(\phi)$ forms a torsion subgroup of $E_1$. An isogeny can be broken down into simpler components. ``Specifically, any isogeny of degree greater than one can be written as a composition of isogenies of prime degree over the algebraic closure of $\mathbb{F}_q$".

\begin{theorem}\cite{Waterhouse1969}
Suppose $E_1$ is an elliptic curve over a field $\mathbb{F}_{q} $ and $G$ be its torsion subgroup. Then there exists a unique elliptic curve $E_2$ over a field $\mathbb{F}_{q}$ and a separable isogeny $\phi: E_1 \rightarrow E_2 $ with $ker(\phi)=G$ such that $E_2 \cong E_1/G $ which can be calculated using Vélu's formula. 
\end{theorem}
The separable isogeny $\phi$ of degree $d= \prod_{i=1}^n {p_i^{e_i}} $, $\phi$ can be broken down as a composition of $e_i$ isogenies of degree $p_i$, for $i=1,2, \cdots, n.$\\

 \subsection{Endomorphism Ring} 
  An isogeny from the curve $E(\mathbb{F}_{q})$ to itself is called an endomorphism. Consider the isogeny $\phi$ as $\phi :E \longrightarrow E$. The collection of all such endomorphisms of $E$, together with the zero maps, forms a ring under the operations \textit{ pointwise addition} and \textit{mapping composition} called the endomorphism ring of $E$, which is represented by $End(E)$. An elliptic curve's endomorphism ring $End(E)$ can be identified with an order in quaternion algebra or  in an imaginary quadratic field. In the former case, the underlying elliptic curve $E$ is said to be supersingular, while in the latter case, it is called ordinary.

 \subsection{Ideal Class Group}

 The elliptic curves $E_1(\mathbb{F}_{q})$ and $E_2(\mathbb{F}_{q})$ are  isogenous if there exists an isogeny $\phi : E_1(\mathbb{F}_{q}) \rightarrow E_2(\mathbb{F}_{q}) $. Tate’s theorem asserts that two elliptic curves over $\mathbb{F}_{q} $ are isogenous if and only if they have the same cardinality i.e., $\#E_1(\mathbb{F}_{q})=\#E_2(\mathbb{F}_{q})$. Since every isogeny has a corresponding dual isogeny, the relation of being isogenous defines an equivalence class over the set of $\overline{\mathbb{F}_{q} }$-isomorphism classes of elliptic curves. These equivalence classes are referred to as isogeny classes. Elliptic curves included in these isogeny classes are all ordinary or all supersingular. In this manuscript we are interested only in isogeny classes that contain supersingular elliptic curves. 

 The $\ell$-torsion group of an elliptic curve $E$, denoted by $E[\ell]$ is defined as
 $$E[\ell]=\{ P \in E(\overline{\mathbb{F}_{q} }) ~ \vert ~\ell P = \O _E \}$$
 
 If $\phi : E_1 \longrightarrow E_2 $ is an isogeny between two elliptic curves then, there exists a corresponding dual $\phi^{'} : E_2 \longrightarrow E_1 $ such that $\phi \circ \phi^{'}=[deg \phi ]$, where $[\ell ]$ is a multiplication-by-scalar-$\ell$ map, which happens to be an endomorphism on $E$. 
Recall, $\mathbb{Q}(\sqrt{-p} )$ is a quadratic field. The class group , denoted by $\O$, is defined as
$\O = \mathbb{Z}[\sqrt{-p} ]=\{ a+b\sqrt{-p} ~\vert~a, b \in \mathbb{Z} \}$, an order or subring of the field $\mathbb{Q}(\sqrt{-p} )$. Hence, $\O$ can be seen as a finitely-generated $\mathbb{Z}$-module containing a basis of $\mathbb{Q}(\sqrt{-p} )$ as a $\mathbb{Q}$-vector space. It is known that 
$\O \cong End_p(E)$, where $End_p(E)$ is the restriction of $ End(E) $ defined over $\mathbb{F}_p$.

\subsection{Class Group Action}
\begin{theorem}\cite{Waterhouse1969}
Consider $\mathcal{O}$ as an order of an $\Q(\sqrt{-p})$ and $E(\mathbb{F}_{q})$ be an elliptic curve. Suppose $\Ell$ denotes the Fp-isomorphism class of supersingular curves, then the action of $\C$ on $\Ell$, defined by

$$ \C \times \Ell\longrightarrow \Ell $$
$$( [\mathfrak{a}], E ) \longrightarrow E/E[\mathfrak{a}]$$
\noindent is free and transitive group action for an ideal $\a \in \mathcal O$.
\end{theorem}
The construction of the $\C$ is explained by Beullens et al. in \cite{beullens2019csi}. The order ${\mathcal O}$ is subring of $End(E)$ and it is isomorphic to $ \Z[\sqrt{-p}]$, where $p$ is a prime of the form $p = 4l_1l_2\cdots l_{n-1}$,  and $l_i$’s are small distinct odd primes with $n = 74, l_1 = 3, l_{73} = 373$, and $l_{74} = 587$.
 $\C$ forms a cyclic group that is generated by $\g = \langle 3, \pi- 1\rangle$. The cardinality of $\C$, which is also known as the class number, is denoted by $N$. Thus, we can consider $\C$ to be $\Z_N$ for simplicity. 
 
\begin{definition}[Perfect Completeness]
A $\Sigma$-protocol is said to have \textbf{perfect completeness} if, when an honest prover and an honest verifier execute the protocol, the verifier will always accept the proof with probability 1.
\end{definition}


\begin{definition}[Special Soundness]
A $\Sigma$-protocol is said to have \textbf{special soundness} if there exists a polynomial-time extractor $\text{Ext}$ such that, for two valid transcripts
\[
\tau_1 = (\mathsf{com}, \mathsf{ch}_1, \mathsf{rsp}_1) \quad \text{and} \quad \tau_2 = (\mathsf{com}, \mathsf{ch}_2, \mathsf{rsp}_2)
\]
, a identical commitment $\mathsf{com}$ and a identical public input $X$, where $\mathsf{ch}_1 \neq \mathsf{ch}_2$, the extractor can compute a valid witness $W$ for the statement associated with $X$.
\end{definition}

\begin{definition}[Honest Verifier Zero-Knowledge]
A $\Sigma$-protocol is said to satisfy \textbf{HVZK} for a given  public parameter $X$, when a polynomial-time simulator $\mathsf{Sim}$, able to produces a transcript $\tau = (\mathsf{com}, \mathsf{ch}, \mathsf{rsp})$ such that the distribution of $\tau$ is computationally indistinguishable to the distribution of transcripts generated during an real execution of the protocol with input $X$.
\end{definition}

\section{Identity-Based Identification (IBID)}\label{sectionIBID}

Through out this section, we are going to discuss our proposed modification of the existing identity-based identification scheme of \cite{shaw2021identification} that relies on a \( T_0 \times S_1 \) matrix representation to instead work with a vector of dimension \( n \), where \( n \) represents the security parameter \cite{katsumata2024csi}. The primary advantage of this transformation lies in the reduced key size and simplified computations, which enhance efficiency while maintaining security by relying on the quadratic twist. If \( A = [\g^a] * E_0 \), where \( a \in \mathbb{Z}_N \) is hidden, then the quadratic twist \( A^{-1}= [\g^{-a}] * E_0 \), can be derived easily. Additionally, we will adopt the (super) exceptional set (CSI-SharK \cite{atapoor2023csi}) to achieve the singular nature of the master secret key. 
\begin{definition}[(Super-)Exceptional set]
An \emph{exceptional set} modulo $N$ is a set $\mathbf c = \{c_1, \ldots, c_{n}\} \subseteq \mathbb{Z}_N$, where the pairwise difference $c_i - c_j$ of all elements $c_i \neq c_j$ is invertible modulo $N$. A \emph{(Super-)Exceptional  set} modulo $N$ is an exceptional set ${\mathbf c} = {c_1, \ldots, c_{n}}$ in which every sum $c_i + c_j$ (including the case $i = j$) has its inverse in modulo $N$.
\end{definition}
As shown in \cite{baghery2021isogeny}, retrieving the secret key using its corresponding public key relies on the hardness of a well-established mathematical problem.
\begin{definition}[($c_1, c_2, \cdots, c_{n}$)-\textbf{Vectorization Problem with Auxiliary Inputs ($\mathbf{c}$-VPwAI)}]\label{exceptionalset}
 For the curve $E_0 \in \Ell$ and the pairs $(c_i, [\g^{c_i x}] E_0)_{i=1}^{n}$, where $\mathbf{c} = \{c_1 = 1, c_2, \ldots, c_{n}\}$ is an exceptional set, compute $x \in \mathbb{Z}_N$.
\end{definition}

Suppose the public parameter is constituted of \( (p, N, E_0) \)  where $p$ is prime, $N$ is the order of the group, and $E_0$ is a distinguished element. Consider the ideal class group \( \mathcal{C}\ell(\mathcal{O}) \) generated by the element \( \g \) . Throughout this work, we presume that these parameters are accessible to every algorithm under consideration. We'll take a hash function \(\hash : \{0,1\}^* \to \{-1,1\}^n \) modeled as a random oracle in the security proof.

\begin{definition}
The ID-based identification protocol $\mathsf{IBID}$ consists of $(\mathsf{IBID.Setup}, \mathsf{IBID.Extract}, \mathsf{Identification \ Protocol})$.
\begin{itemize}
    \item  \textbf{$\mathsf{IBID.Setup}$:} Given a security parameter \( 1^n \), it chooses the base elliptic curve $E_0: y^2=x^3+x\in \Ell$, and a public (super) exceptional set $\mathbf{c}=\{c_1=1,c_2,\cdots, c_n\}$. It also samples $s\xleftarrow{\$} \mathbb{Z}_N$, and set $\mathbf E=[\g^{s\cdot \mathbf c}]*E_0$.  Finally it outputs the public parameter $\param$ be $(p, N, E_0, \g, \hash, \mathbf{c}, \mathbf E)$ and master secret key  $\sk=\{s\}$.

\item  \textbf{$\mathsf{IBID.Extract}$:} On input the security parameter $(\param, \sk, \id) $, this algorithm first samples $\mathbf  r \xleftarrow{\$}  \Z_N^n$, and computes $\mathbf R=[\g^{\mathbf r}]*E_0$. Then it computes $\{ \mathbf u\}=\hash (\id \concat \mathbf R )$, and secret key values of the prover $\mathbf x =\mathbf r-s\cdot \mathbf c \odot \mathbf u \ (\bmod \  N)$. It then  the outputs user private key $\mathsf{usk_{id}}=(\{\mathbf u, \mathbf x\}$)

\item $\mathsf{Identification \ Protocol}:$ This is described in Fig. \ref{fig:Id-protocol} which is an interactive session between the prover $P$ with input $(\param$, $\mathsf{usk_{id}})$ and the verifier $V$ with input $(\param$, $\id)$. On completion of this protocol, the $V$ outputs $1$, signifying ``accept"; otherwise, $0$.
\end{itemize}

\begin{figure}[h!]
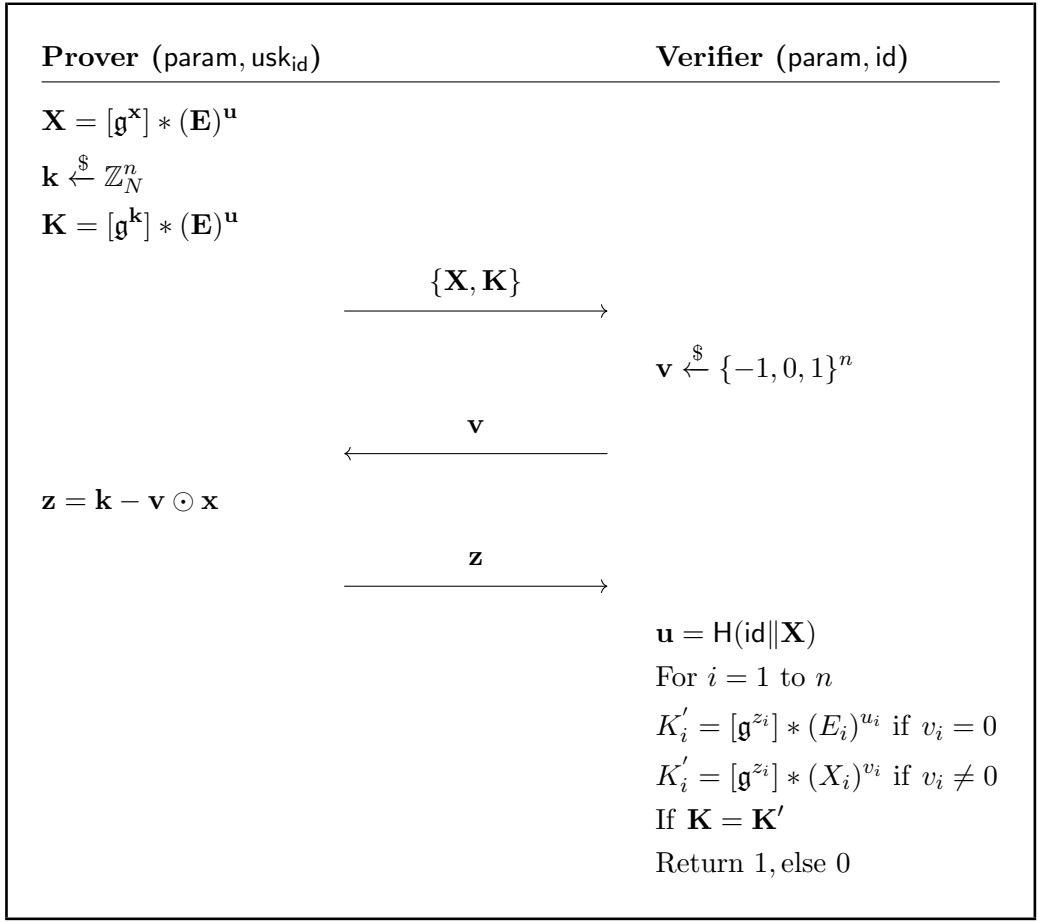

\begin{mdframed}[roundcorner=10pt, innerleftmargin=10pt, innerrightmargin=10pt, innertopmargin=10pt, innerbottommargin=10pt, linewidth=1pt]
\pseudocodeblock{
\textbf{Prover ($\param, \mathsf{usk_{id}}$)} \< \< \textbf{Verifier ($\param, \id$)}\\
[0.1 \baselineskip] [\hline]
\<\<\\[-0.5\baselineskip]
\mathbf{X} = [\g^{\mathbf x}]*(\mathbf E)^{\mathbf u} \\
\mathbf k \xleftarrow{\$}  \Z_N^n \\
\mathbf K = [\g^{\mathbf k}]*(\mathbf E)^{\mathbf  u} \\
\< \sendmessageright*{\{\mathbf X, \mathbf K\} } \<\\
\< \< \mathbf v \xleftarrow{\$}  \{-1,0,1\}^n\\
\< \sendmessageleft*{\mathbf v }\<\\
\mathbf z = \mathbf k-\mathbf v\odot \mathbf x \\
\< \sendmessageright*{\mathbf z}\<\\
\< \<  \mathbf u =\hash (\id \concat \mathbf X)\\
\< \<  {\rm For \ } i=1  {\rm \ to \ } n\\
\< \< K_{i}^{'}=[\g^{z_i}]*(E_{i})^{u_{i}} {\rm \ if \ }v_i=0 \\
\< \< K_{i}^{'}=[\g^{z_i}]*(X_{i})^{v_{i}} {\rm \ if \ }v_i\neq 0 \\
\< \<{\rm If\ } \mathbf K =  \mathbf K'  \\
\< \< {\rm Return\ } 1, {\rm else \ } 0 
}
\end{mdframed}
\caption{Flow diagram of identification scheme followed by prover $P$ and verifier $V$.}
\label{fig:Id-protocol}
\end{figure}
\end{definition}

\section{Proposed construction of our Identity-based Blind Signature(IBBS)}\label{sectionIBBS}
\subsection{Base sigma protocol}
We build our scheme from a sigma protocol that demonstrates the signer's knowledge of at least one of two private keys $(x_0,x_1)$ associated with the public keys 
\[
X = ( X_0,  X_1) = ([\g^{ x_0}]*(E_0), [\g^{ x_1}]*(E_0)).
\] 

\begin{figure}[h!]
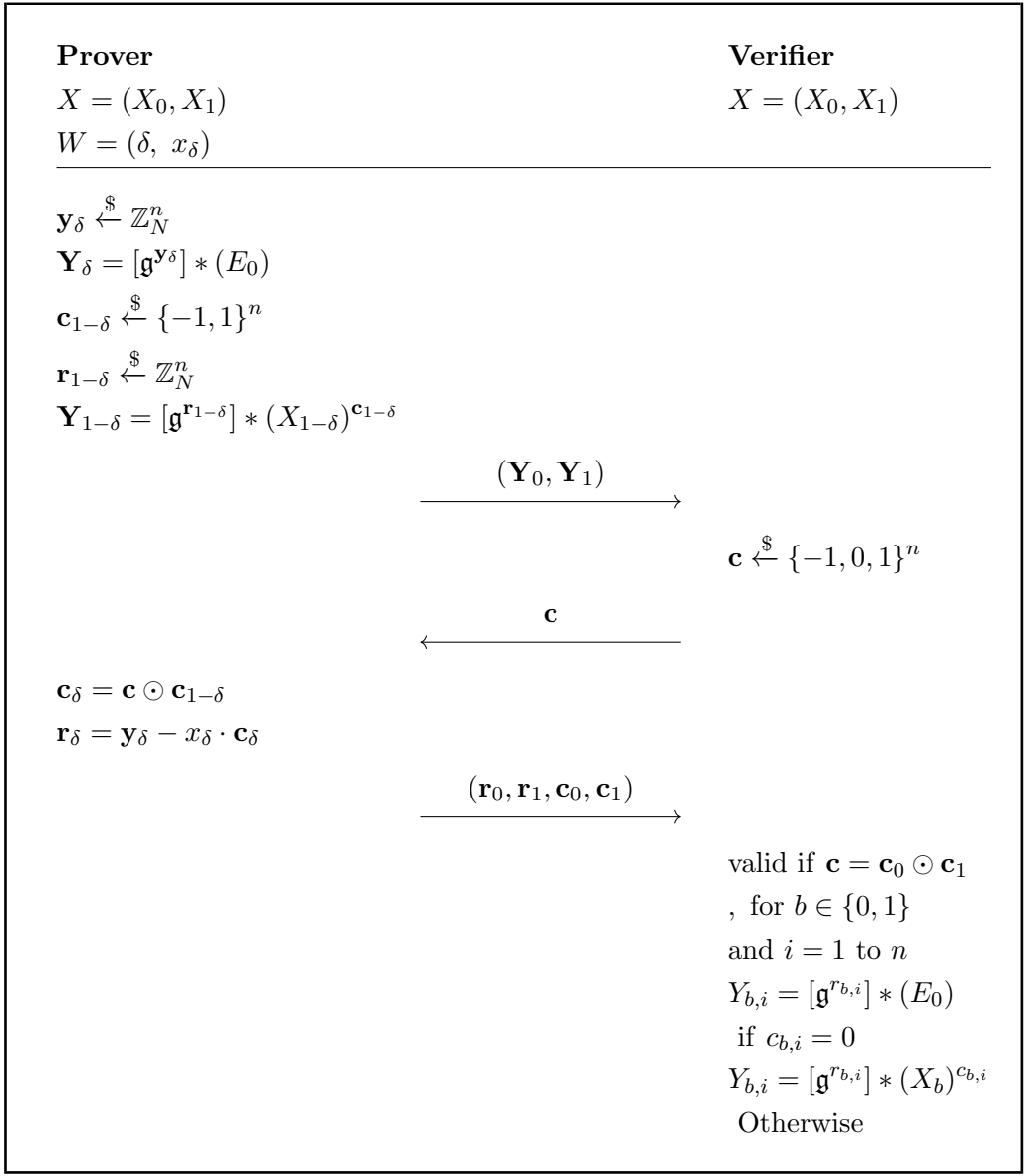

\begin{mdframed}[roundcorner=10pt, innerleftmargin=10pt, innerrightmargin=2 pt, innertopmargin=10pt, innerbottommargin=8 pt, linewidth=1pt]
\pseudocodeblock{
\textbf{Prover} \< \< \textbf{Verifier}\\
 X=( X_0,  X_1)  \< \< X=( X_0, X_1) \\
W=(\delta,\ x_\delta) 
\\[0.1 \baselineskip] [\hline]
\<\<\\[-0.5\baselineskip]
\mathbf y_{\delta} \xleftarrow{\$}  \Z_N^n \\
\mathbf Y_{\delta} = [\g^{\mathbf y_{\delta}}]*(E_0) \\
\mathbf c_{1-\delta} \xleftarrow{\$}  \{-1,1\}^n \\
\mathbf r_{1-\delta} \xleftarrow{\$}    \Z_N^n \\
\mathbf Y_{1-\delta} = [\g^{\mathbf r_{1-\delta}}]*(X_{1-\delta})^{\mathbf c_{1-\delta}} \\
\< \sendmessageright*{(\mathbf Y_0, \mathbf Y_1)}\<\\
\< \< \mathbf c \xleftarrow{\$}  \{-1,0,1\}^n\\
\< \sendmessageleft*{\mathbf c }\<\\
\mathbf c_{\delta} = \mathbf c\odot \mathbf c_{1-\delta} \\
\mathbf r_{\delta} = \mathbf y_{\delta}-x_{\delta}\cdot \mathbf c_{\delta} \\
\< \sendmessageright*{(\mathbf r_0, \mathbf r_1, \mathbf c_0, \mathbf c_1)}\<\\
\< \< {\rm valid\ if\ } \mathbf  c=\mathbf c_0\odot \mathbf c_1  \\
\< \< {\rm ,\ for\ } b\in\{0,1\}\\
\< \< {\rm and \ } i=1  {\rm \ to \ } n\\
\< \< Y_{b,i}=[\g^{r_{b,i}}]*(E_{0}) \\
\< \< {\rm \ if \ }c_{b,i}= 0 \\
\< \< Y_{b,i}=[\g^{r_{b,i}}]*(X_{b})^{c_{b,i}} \\
\< \<{\rm \ Otherwise \ }
}
\end{mdframed}
\caption{The sigma protocol that serves as the basis for our IBBS scheme.}
\label{fig:base-or-sigma-protocol}
\end{figure}

This sigma protocol, illustrated in Figure \ref{fig:base-or-sigma-protocol}, follows a standard isogeny-based approach, similar to \cite{Katsumata2024group}. However, we note that our protocol includes $0$ in the challenge space, which differs from \cite{Katsumata2024group}. This adjustment prevents the challenge space from forming a multiplicative subgroup of $\mathbb{Z}_N^{\times}$. While this introduces a trade-off, it enables the development of our IBBS scheme, where both the master public key and user public keys are used for verification. The correctness of the protocol can be readily confirmed with a simple check.

\subsubsection{HVZK}

For a known challenge $\mathbf{c}$, the zero-knowledge simulator $\mathsf{Sim}$ chooses random vectors $ (\mathbf{c}_0, \mathbf{c}_1) \xleftarrow{\$} (\{-1, 0, 1\}^n)^2 $ and $(\mathbf{r}_0, \mathbf{r}_1) \xleftarrow{\$} (\mathbb{Z}_N^{n})^2,
$ subject to the condition $\mathbf{c}_0 \odot \mathbf{c}_1 = \mathbf{c}$. And, for every $(b, i) \in \{0,1\} \times [1\!:\!n]$, it computes
\[
Y_{b,i} = 
\begin{cases}
[\mathbf{g}^{r_{b,i}}] \ast (X_i)^{c_{b,i}}, & \text{if } c_{b,i} \neq 0, \\
[\mathbf{g}^{r_{b,i}}] \ast (E_0), & \text{otherwise}.
\end{cases}
\]
Finally, it outputs the simulated transcript $ \left(\mathbf{Y}_b,\, \mathbf{c},\, (\mathbf{r}_b, \mathbf{c}_b)\right)_{b \in \{0,1\}} $. Since $\mathbf{c}_b$ determines a bijection between $\mathbf{r}_b$ and $\mathbf{Y}_b$, the distribution of resulting transcript is indistinguishable from that of a actual interaction.

We note that although a larger challenge space $\mathcal{C} = \{-1, 0, 1\}^n$ may slightly increase computational cost, it also reduces the soundness error, and supports special soundness by making multiple challenge collisions less likely--a property we will analyze in the next subsection.

\subsubsection{Special Soundness}
Consider $((\mathbf{Y}_0, \mathbf{Y}_1), \mathbf{c}, (\mathbf{r}_0, \mathbf{r}_1, \mathbf{c}_0, \mathbf{c}_1))$, and $ ((\mathbf{Y}_0, \mathbf{Y}_1), \mathbf{c}', (\mathbf{r}_0', \mathbf{r}_1', \mathbf{c}_0', \mathbf{c}_1'))$
as  valid transcripts such that \( \mathbf{c} \neq \mathbf{c}' \). Inequality \( \mathbf{c} \neq \mathbf{c}' \), implies that at least one of the conditions \( \mathbf{c}_0 \neq \mathbf{c}_0' \) and \( \mathbf{c}_1 \neq \mathbf{c}_1' \) is true. We may assume for simplicity that \( \mathbf{c}_{0,1} \neq \mathbf{c}_{0,1}' \), in accordance with \( \mathbf{c}_{0,1} \in \{-1, 0, 1\} \) is index-1 element of \( \mathbf{c}_0 \) and \(\mathbf{c}_{0,1}' \in \{-1, 0, 1\}\) is index-1 element of \( \mathbf{c}_0' \).  

The extractor \( \mathsf{Ext} \) subsequently produces a witness  
\[
(0, x_0 = \frac{r_{0,1} - r_{0,1}'}{c_{0,1} - c_{0,1}'}),
\]  
where \( r_{0,1}, r_{0,1}' \in \mathbb{Z}_N \) are the first elements of \( r_0 \) and \( r_0' \). Note that since \( p \equiv 3 \pmod{4} \), \( N \) is odd, so \( c_{0,1} - c_{0,1}' \in \{-2, -1, 1, 2\} \) is invertible modulo \( N \).

\subsection{Structure of our Identity-based Blind Signature}

The foundation of our proposed scheme is built upon the base sigma protocol Figure \ref{fig:base-or-sigma-protocol}, similar to \cite{katsumata2024csi}. Consider \( (p, N, E_0) \) as public parameter, where $p$ is prime, $N$ denotes the group order, and $E_0$ is the starting curve. We presume that all algorithms have access to these parameters. Consider \(\hash : \{0,1\}^* \to \{-1,1\}^n \) and \(\tilde{\hash}:\{0,1\}^* \to \{-1,0,1\}^n \) as a hash function modeled in random oracle for the security analysis.  $\mathsf{IBBS.Setup}$ and $\mathsf{IBBS.Extract}$ of the protocol are executed by $\mathsf{KGC}$.\\

\noindent \textbf{$\mathsf{IBBS.Setup}$:} Initialized the system parameter according to the security parameter \(1^n\),
\begin{enumerate}
     \item Choose a set of $n$ small prime ideals $\{l_i\}$, a starting curve $E_0:y^2=x^3+x $, and (Super-)exceptional set $\mathbf{c}=\{c_1=1,c_2,\cdots, c_n\}$. 
    \item It chooses master secret key $\{s_0,s_1\}\xleftarrow{\$}  \Z_N^2$, and computes the corresponding elliptic curves (master public key) ${\mathbf E}_b = [\g^{{s_b\cdot \mathbf c}}]*E_0$ for $b\in \{0,1\}$.
    \item It generates the corresponding public parameter $\param$ be $(p, N, E_0, \g, \hash, \mathbf{c}, \mathbf E_b)$ and master secret key  $\sk=\{s_0,s_1\}$.
\end{enumerate}

\noindent \textbf{$\mathsf{IBBS.Extract}$:} Given  ($\param, \sk, \id$) act as follows,
\begin{enumerate}
 \item The $\mathsf{KGC}$ parses $\param$, $\sk$, and utilizes the user’s identity ($\id$) to compute secret values. It first samples ${\mathbf r}_b \xleftarrow{\$}  \Z_N^n$, and computes the corresponding elliptic curves as $\mathbf X_b=[\g^{\mathbf r_b}]*E_0$. 
 \item The hash function $\hash$ processes the identity of the user, system parameters, along with the generated curves to produce hash values as $\mathbf u_b =\hash (\id \concat \mathbf X_b)$. 
 \item The secret key values of the signer are computed as $\mathbf x_b =\mathbf r_b-s_b \cdot \mathbf u_b \odot \mathbf c$.
 \item A bit $\delta \xleftarrow{\$}  \{0,1\}$ is sampled, and the user secret key is returned as $\mathsf{usk_{id}}=(\delta, \mathbf x_\delta)$, along with the public key $\mathsf{pk_{id}}=(\mathbf X_0, \mathbf X_1)$.
\end{enumerate}

\vspace{0.5 cm}
\noindent \textbf{$\mathsf{IBBS.S_1}$:} Given  $(\param,\mathsf{pk_{id}}, \mathsf{usk_{id}})$, to initiate the signing of a message, the signer executes the following steps:
\begin{enumerate}
 \item First, $\mathbf y_\delta  \xleftarrow{\$}  \Z_N^n$, and then 
 computes $\mathbf Y_\delta=[\g^{\mathbf y_\delta}]*(\mathbf E_\delta)^{\mathbf u_\delta}$. Also computes $\mathbf u_b =\hash (\id \concat \mathbf X_b)$ for $b\in\{0,1\}$. 
 \item Next, $(\tilde{\mathbf c}_{1-\delta}, \mathbf r^*_{1-\delta}) \xleftarrow{\$}  \{-1,1\}^n\times \Z_N^n$.
 \item It computes $\mathbf c^*_{1-\delta}=\tilde{\mathbf c}_{1-\delta}\odot \mathbf u_{1-\delta}$, and $\mathbf Y_\mathbf {1-\delta}=[\g^{\mathbf r^*_{1-\delta}}]*(\mathbf X_{1-\delta})^{\mathbf c^*_{1-\delta}}$
 \item It generates first-sender message $\mathsf{\rho_{S,1}}=(\mathbf Y_{0}, \mathbf Y_{1})$, and $\mathsf{State}_S=(\mathbf y_{\delta}, \mathbf c^*_{1-\delta}, \mathbf r^*_{1-\delta})$ 
\end{enumerate}

\vspace{0.5 cm}
\noindent \textbf{$\mathsf{IBBS.U_1}$:} Given  $(\param, \mathsf{\rho_{S,1}}, \m)$, the user executes the following steps:
\begin{enumerate}
 \item Parse $(\mathbf Y_{0}, \mathbf Y_{1}) \gets \mathsf{\rho_{S1}}$ 
 \item It samples $(\mathbf v_b, \mathbf w_b) \xleftarrow{\$}  \{-1,1\}^n\times \Z_N^n $, and then computes $\mathbf Z_b=[\g^{\mathbf w_b}]*(\mathbf Y_{b})^{\mathbf v_b}$, for $b\in\{0,1\}$. 
 \item Calculates $\mathbf c = \tilde \hash (\mathbf Z_0 \concat \mathbf Z_1 \concat \m) $
 \item Calculates $\mathbf c^{*} = \mathbf c \odot \mathbf v_0 \odot \mathbf v_1 $
 \item It outputs user message $\mathsf{\rho_{U}}=\mathbf c^{*}$, and user state $\mathsf{State}_U=(\mathbf v_b, \mathbf w_b)$
\end{enumerate}

\vspace{0.5 cm}
\noindent \textbf{$\mathsf{IBBS.S_2}$:} Given $(\mathsf{State}_S, \mathsf{\rho_{U}})$, the signer signs the blinded message by executing the following steps:
\begin{enumerate}
 \item It parse $(\mathbf c^{*})\gets \mathsf{\rho_{U}}$, and $(\mathbf y_{\delta},\mathbf c^*_{1-\delta}, r^*_{1-\delta})\gets \mathsf{State}_S$
 \item Next it computes $\mathbf c^*_{\delta}=\mathbf c^{*}\odot \mathbf c^*_{1-\delta}$, and $\mathbf r^*_{\delta}=\mathbf y_{\delta} -\mathbf x_{\delta}\odot \mathbf c^*_{\delta}$. 
 \item It outputs second-sender message $\mathsf{\rho_{S,2}}=(\mathbf c^*_{0}, \mathbf c^*_{1}, \mathbf r^*_{0}, \mathbf r^*_{1})$. 
\end{enumerate}

\vspace{0.5 cm}
\noindent \textbf{$\mathsf{IBBS.U_2}$:} Given  $(\mathsf{pk_{id}} ,\mathsf{\rho_{U}}, \mathsf{\rho_{S,2}},  \mathsf{State}_U)$, following steps will be executed.
\begin{enumerate}
 \item It parse $(\mathbf X_{0}, \mathbf X_{1})\gets \mathsf{pk_{id}}$, $(\mathbf c^*_{0}, \mathbf c^*_{1}, \mathbf r^*_{0}, \mathbf r^*_{1}) \gets \mathsf{\rho_{S2}}$, $\mathbf c^{*}\gets \mathsf{\rho_{U1}}$, and $(\mathbf v_b, \mathbf w_b)\gets \mathsf{State}_U$.
 \item It computes $\tilde{\mathbf c}_b=\mathbf c^*_{b}\odot \mathbf v_b$, and $\tilde{\mathbf r}^*_{b}= \mathbf w_b + \mathbf r^*_{b}\odot \mathbf v_b$ for $b\in \{0,1\}$. 
  \item For $b\in\{0,1\}$, compute $\hash (\id \concat \mathbf X_b)$, and parse it into $\tilde{\mathbf u}_b$. 
  \item Compute the response values $\tilde{\mathbf Z}_b$:
    \begin{itemize}
        \item If $\tilde{c}_{b,i} = 0$, then $\tilde{Z}_{b,i}=[\g^{\tilde{r}_{b,i}}]*(E_{b,i})^{\tilde{ u}_{b,i}}$
        \item Otherwise, $\tilde{Z}_{b,i}=[\g^{\tilde{r}_{b,i}}]*(X_{b,i})^{\tilde{c}_{b,i}}$
    \end{itemize}
 \item Hash the computed curves, and message to obtain ${\hat {\mathbf c}}= \tilde \hash (\tilde{ \mathbf Z}_0\concat \tilde{\mathbf Z}_1\concat \m)$
 \item Check if $\tilde{\mathbf c}_0 \odot \tilde{\mathbf c}_1 = {\hat {\mathbf c}}$. If it satisfies, then the generated signature is ${\sigma}=(\tilde{\mathbf c}_{b}, \tilde{\mathbf r}_{b})$, for $b\in \{ 0,1 \}$; otherwise a $\perp$.
  \end{enumerate}

\vspace{0.5 cm}
\noindent \textbf{$\mathsf{IBBS.Verify}$:} Given $(\param,\mathsf{pk_{id}}, \sigma, \m)$, the verifier executes the following steps:
\begin{enumerate}
    \item It first parses $(\mathbf X_{0}, \mathbf X_{1})\gets \mathsf{pk_{id}}$, and $(\tilde{\mathbf c}_{b}, \tilde{\mathbf r}_{b}) \gets \sigma$ .
    \item For $b\in\{0,1\}$, compute $\hash (\id \concat \mathbf X_b)$, and parse it into $\tilde{\mathbf u}_b$. 
    \item Compute the response values $\tilde{\mathbf Z}_b$:
    \begin{itemize}
        \item If $\tilde{c}_{b,i} = 0$, then $\tilde{Z}_{b,i} =[\g^{\tilde{r}_{b,i} }]*(E_{b,i})^{\tilde{u}_{b,i}}$
        \item Otherwise, $\tilde{Z}_{b,i}=[\g^{\tilde{r}_{b,i}}]*(X_{b,i})^{\tilde{c}_{b,i}}$
    \end{itemize}
 \item Finally, it hashes the computed curves and message to obtain ${\hat {\mathbf c}}= \tilde \hash (\tilde{ \mathbf Z}_0\concat \tilde{\mathbf Z}_1\concat \m)$
 \item Check if ${\hat {\mathbf c}}=\tilde{\mathbf c}_0 \odot \tilde{\mathbf c}_1$. If true, the signature is accepted as valid; otherwise, it is rejected.
 \end{enumerate}

{\color{blue}

}
\section{Security Analysis}\label{sectionSecurity}
 Now, we are going to discuss rigorous security analysis of the {Identity-Based Identification scheme ($\mathsf{IBID}$)}, and {Identity-Based Blind Signature scheme ($\mathsf{IBBS}$)} rely on the {CSIDH} framework. The proof focuses on the correctness, efficiency, and impersonation resistance for the $\mathsf{IBID}$ scheme and unforgeability and blindness properties of the scheme, demonstrating its security under the Existential Unforgeability under Chosen Message Attack (EUF-CMA) model. The security of our schemes relies on the problems defined below:
\begin{definition}[\textbf{Group action inverse problem (GAIP)}]\label{def:GAIP} Consider the  two isogenous supersingular elliptic curve $(E_0, E_A) \in \Ell$, then the GAIP problem is finding an element $[\mathfrak{a}] \in \C$ such that $E_A = [\g^a]E_0$ .
\end{definition}

\begin{definition}[\textbf{Multi-target Group action inverse problem (MT-GAIP)}]\label{def:MTGAIP} For $k-isogenous$ supersingular elliptic curves $E_0, E_1, E_2,\cdots,E_k$ find an an element $[\mathfrak{a}] \in \C$ such that $E_i = [\mathfrak{\g^a}]E_j$ with $i \neq j$.
\end{definition}


\subsection{Security analysis of ${\mathsf{IBID}}$}
The cornerstone of our construction lies in the identification scheme of \cite{shaw2021identification} based on identity-based cryptography (IBC). Notably, their security is analyzed under the framework of Impersonation under Passive Attacks IMP-PA\cite{shaw2021identification}, ensuring rigorous evaluation against potential threats. To establish a robust foundation, we start with the security model of the ${\mathsf{IBID}}$.
\begin{enumerate}
    \item \textbf{Correctness:} To ensure the correctness of our ${\mathsf{IBID}}$ scheme, it is sufficient to show that the Prover and the Verifier adhere to the construction in Figure \ref{fig:Id-protocol}. First, note that if $v_i\neq 0$, then $K^{'}_{i}=[\g^{z_i}]*(X_{i})^{v_{i}}=[\g^{k_i-v_i*x_i}]*(X_{i})^{v_{i}}=K_i$. Similarly $K^{'}_{i}=K_i$, if $v_i=0$.
    \item \textbf{Efficiency}
    The efficiency of the proposed IBID protocol is comparable to the original version presented in \cite{shaw2021identification}. However, in contrast to  CSI-SharK protocol, which depends on the master key composed of $S_0$ elements in $\mathbb{Z}_N$, our schemes simplifies the setup by using just one element from  $\mathbb{Z}_N$ as master secret key. Moreover, we leverage a quadratic twist, which we can compute efficiently.
    \item \textbf{Impersonation Resistance:}
    To show this security proof, we follow the framework of \cite{shaw2021identification}, which establishes security for identity-based identification protocols. We adapt their methodology with two key modifications:

\begin{enumerate}
    \item \textbf{Efficiency Improvement via Vectorization}:
    Instead of the original $T_0 \times S_1$ matrix representation, we adopt an $n$-dimensional vector structure. This reduces computational overhead while preserving security, as scalar operations via the quadratic twist remain hard under the same assumptions.

    \item \textbf{Key Structure via (Super) Exceptional Sets}:
    To ensure the master key has singular size, we sample it from a (super) exceptional set modulo $N$. By Definition~\ref{exceptionalset}, this guarantees invertibility of pairwise sums/differences, preventing degeneracy in key generation.
\end{enumerate}

These changes necessitate adjustments to the above $\mathsf{IBID}$ scheme. the security holds because:
\begin{enumerate}
    \item The quadratic twist preserves the hardness of the underlying problem, and
    \item Exceptional sets ensure non-trivial linear independence in key relations.
\end{enumerate}

Thus, any adversary $\mathcal{A}$ breaking our scheme implies an attack on \cite{shaw2021identification}'s construction, up to a negligible loss in the reduction.

\begin{theorem}
Consider $\mathcal{A}$ as an impersonator (under passive attack) capable of retrieving the private key of the $\mathsf{IBID}$ (IMP-PA security \cite{shaw2021identification}). Then, we can construct an adversary $\mathcal{B}$ that retrieves the private key of the signature protocol $\mathsf{CSI-FiSh}$ against existential unforgeability under an adaptive chosen message attack. 
\end{theorem} 

\end{enumerate}



\subsection{Security analysis of ${\mathsf{IBBS}}$}
Throughout the section, we are going to discuss the game-based analysis to demonstrate the EUF-CMA security of the IBBS protocol. Using the Schnorr-like identification scheme for effective signature generation, the proof takes advantage of the CSIDH's hard problem, like GAIP  and the MT-GAIP, which are utilized in CsiIBS. By utilizing the CSIDH group action and the CSI-Otter quadratic twist optimizations, the IBBS scheme combines the CsiIBS and CSI-Otter frameworks. Among the security attributes are:

\begin{itemize}
\item \textbf{Correctness:} To ensure correctness, it is sufficient to show that signer and the user correctly execute the ${\mathsf{IBBS}}$ scheme. First, observe that since ${\mathbf r}_{b}={\mathbf x}_{b}+s_b\cdot {\mathbf u}_{b}\odot \mathbf c$, it follows that $\mathbf X_b=[\g^{\mathbf r_b}]*E_0=[\g^{\mathbf x_b}]*(\mathbf E_b)^{\mathbf u_b}$. Additionally, we have ${\mathbf c}=\mathbf c^{*}\odot \mathbf v_0\odot \mathbf v_1=\mathbf c^*_{0}\odot \mathbf c^*_{1}\odot \mathbf v_0\odot \mathbf v_1=\tilde{\mathbf c}_0 \odot \tilde{\mathbf c}_1$. Here, we use the property that for any $v\in\{-1,1\}$, it holds that $v\odot v=1$. Now, when $\tilde{c}_{b,i}\neq 0$, we derive the following sequence of equalities: \begin{eqnarray*}
    [\g ^{\tilde{r}_{b,i}}]*(X_{b,i})^{\tilde{c}_{b,i}}
    & = &  [\g ^{\tilde r_{b,i} +{x_{b,i}\odot \tilde{c}_{b,i} }}]*(E_{b,i})^{u_{b,i}}\\
    & = & [\g ^{w_b+r^*_{b,i}\odot v_{b,i} +{x_{b,i}\odot c^*_{b,i} \odot v_{b,i}}}]*(E_{b,i})^{u_{b,i}}\\
    & = & [\g ^{w_{b,i}+(r^*_{b,i} +{x_{b,i}\odot c^{*} )\odot v_{b,i}}}]*(E_{b,i})^{u_{b,i}}\\
    & = & [\g ^{w_{b,i}+y_{b,i} \odot v_{b,i}}]*(E_{b,i})^{u_{b,i}}\\
    & = & [\g ^{w_{b,i}}]*(Y_{b,i})^{v_{b,i}}\\
    & = & Z_{b,i}
\end{eqnarray*}
Similarly verification follows when $\tilde{c}_{b,i}=0$.
\item \textbf{Blindness:} 
   Blindness can be established by proving that, given a legitimate public key $\mathsf{pk_{id}}$, corresponding signer messages $\mathsf{\rho_{S_1}}$ and $\mathsf{\rho_{S_2}}$, and a valid signature $\sigma = (\tilde{\mathbf{c}}_b, \tilde{\mathbf{r}}_b)$, one can identify a unique user state $\mathsf{State}_U = (\mathbf{v}_b, \mathbf{w}_b)$, which is pairwise distinct from other states and responsible for the signature. For fixed user (private) key $\mathsf{usk_{id}}=(\mathbf x_0, \mathbf x_1)$, public key $\mathsf{pk_{id}}=(\mathbf X_0, \mathbf X_1)$, $\mathsf{\rho_{S,1}}=(\mathbf Y_{0}, \mathbf Y_{1})$, $\mathsf{\rho_{S,2}}=(\mathbf c^*_{b}, \mathbf r^*_{b})$ for $b\in\{0,1\}$, and signature ${\sigma}=(\tilde{\mathbf c}_{b}, \tilde{\mathbf r}_{b})$, we can derive $\mathbf v_b = \mathbf c_b\odot \mathbf c^*_b$ then, $\mathbf w_b = \tilde{\mathbf r}_b -\mathbf r^*_{b}\odot \mathbf v_b$. The verification equations must hold by completeness.

    This User State is unique, as no other pair \((\mathbf{v}^{'}_b, \mathbf{w}^{'}_b)\) can yield the same \(\sigma\).  If \(\mathbf{v}_b' \neq \mathbf{v}_b\), then \(\mathbf{w}_b'\) would need to compensate, violating the verification equation. Conversely, if \(\mathbf{v}_b' = \mathbf{v}_b\), then \(\mathbf{w}_b'\) must necessarily equal \(\mathbf{w}_b\) to preserve \(\tilde{\mathbf{r}}_b\). The states are also pairwise distinct: for two signatures \(\sigma_1=(\tilde{\mathbf c}^{'}_{b},\tilde{\mathbf r}^{'}_{b})\) and \(\sigma_2=(\tilde{\mathbf c}^{''}_{b},\tilde{\mathbf r}^{''}_{b})\), if \(\tilde{c}^{'}_{b} \neq \tilde{c}^{''}_{b}\), then \(\mathbf{v}^{'}_{b} \neq \mathbf{v}^{''}_{b}\). Otherwise, if \(\tilde{c}^{'}_{b} = \tilde{c}^{''}_{b}\) but \(\tilde{\mathbf r}^{'}_{b} \neq \tilde{\mathbf r}^{''}_{b}\), then \(\mathbf{w}^{'}_{b} \neq \mathbf{w}^{''}_{b}\). Thus, each \(\sigma\) maps to a unique \(\mathsf{State}_U\), proving the signer cannot link signatures to protocol executions and establishing blindness.

    \item \textbf{$\mathsf{EUF\text{-}CMA\text{-}CIDA}$ Security:} 
    
    Consider $\mathcal{A}$ as the adversary having access to the system parameters $\param= \{p, N, E_0, \g, \hash, \mathbf{c}, \mathbf E\}$ and the identity public key $\{\mathbf X_0,\mathbf X_1\}$ of the signer ID. $\mathcal{A}$ attempts to forge a valid message-signature of the signer. We show that our IBBS scheme is secure under attacks using adaptive chosen-message and chosen-identity techniques that render the system existentially unforgeable, or alternatively  $$\mathsf{Adv}^{\mathsf{EUF\text{-}CMA\text{-}CIDA}}_{\mathcal{A},{\mathsf{IBBS}}}(\lambda) \leq \mathsf{negl}(\lambda).$$

    First, we assume that $\mathcal{A}$ is the adversary that performs the chosen-identity attack, i.e., $\mathcal{A}$ queries $\mathsf{IBBS.Extract}$ $q_e$ ($0<q_e<<N$) times with $(\param, \id_i)$ for $i\in\{1,2,\cdots,q_e\}$. $\mathsf{IBBS.Extract}$ reutrns the user private keys $\{\delta^i, \mathbf{x}^i_\delta \}$, and the public keys $(\mathbf X^i_0, \mathbf X^i_1)$ with $\mathcal{A}$. The adversary computes $\mathbf u^i_b=\hash (\id \concat \mathbf X^i_b)$ for $b\in\{0,1\}$. However, since $\hash$ is modeled as a random oracle, $\mathcal{A}$ gains no useful information directly from it. Finally $\mathcal{A}$ collects multiple instances of $(\mathbf X^i_b, \mathbf u^i_b,\mathbf x^i_b)_{b\in\{0,1\}}$, and analyzes the equation \( \mathbf x_b = \mathbf r_b - s_b \cdot \mathbf u_b \odot \mathbf c \). Since \( \mathbf r_b \) is randomly chosen and intractable due to the hardness assumption of GAIP/MT-GAIP (Definitions \ref{def:GAIP}, \ref{def:MTGAIP}), and \( \mathbf u_b \) is deterministically computed from \( \mathbf X^i_b\) and \(\mathsf{id} \), an attack is possible if the adversary \( \mathcal{A} \) can solve for \( s_b \) using multiple queries. However,  $q_e$, the number of queries to the $\mathsf{IBBS.Extract}$ is quite less than $N$, and the chances of this attack are infeasible. 

    On the other hand, the (blind) signature and the verifying scheme closely resemble CSI-Otter \cite{katsumata2024csi}. For any message $\m$, their protocol is proven secure against one-more unforgeability (see Theorem 12 in \cite{katsumata2024csi}) under the hardness assumption of GAIP/MT-GAIP. Since our $\mathsf{IBBS}$ scheme follows a similar structure and relies on the same hardness assumptions, we demonstrate that the scheme remains secure against adaptive chosen-message attacks.

\end{itemize}

\section{Performance Analysis}\label{performance}
Throughout this section, we have analyzed the performance of the introduced design in terms of computation cost, key sizes, and signature sizes. We compare our scheme with existing IBBS schemes.

\indent Applying an \(\ell_i\)-degree isogeny using Vélu’s formulas incurs a cost of \( O(\ell_i) \) field operations. Since each \( \ell_i \) is a small constant (e.g., 3, 5, 7, etc.), this results in an effective cost of \( O(1) \) per isogeny computation. For CSIDH group action, multiple \(\ell_i\)-degree isogeny needs to be calculated; the resulting cost of a single group action is given by $O(n)$. This confirms that the overall number of field operations needed grows linearly with the security parameter \( n \).

In practical implementations, such as CSIDH-512, a complete group action consisting of \( n \) small-degree isogenies is empirically found to require approximately \( 10^4 \) to \( 10^5 \) field operations. In addition to isogeny evaluations, auxiliary operations such as vector arithmetic and hashing are also involved, but they too scale linearly with \( n \). Hence, the total complexity of a single CSIDH group action remains \( O(n) \), demonstrating that the cost scales efficiently with the size of the parameter. Table~\ref{tab:costs} provides the computational cost of various algorithms of the proposed IBBS.

\begin{table}[h!]
  \centering
  \resizebox{\textwidth}{!}{
    \begin{tabular}{|l|l|c|}
      \hline
      \textbf{Algorithm} & \textbf{Dominant Operations} & \textbf{Complexity} \\
      \hline
      \(\mathsf{IBBS.Setup}\)   & \(2n\) group actions (compute \(\mathbf E_b=[g^{s_b\cdot\mathbf c}]\ast E_0\))               & \(O(n^{2})\) \\
      \hline
      \(\mathsf{IBBS.Extract}\) & \(2n\) group actions (derive \(\mathbf X_0,\mathbf X_1\)) + hashing                         & \(O(n^{2})\) \\
      \hline
      \(\mathsf{IBBS.S_1}\)     & \(2n\) group actions (generate \(\mathbf Y_0,\mathbf Y_1\))                                  & \(O(n^{2})\) \\
      \hline
      \(\mathsf{IBBS.U_1}\)     & \(2n\) group actions (form \(\mathbf Z_0,\mathbf Z_1\)) + hashing                            & \(O(n^{2})\) \\
      \hline
      \(\mathsf{IBBS.S_2}\)     & \(2n\) scalar multiplications/additions in \(\mathbb{Z}_N\) (no isogenies)                   & \(O(1)\) \\
      \hline
      \(\mathsf{IBBS.U_2}\)     & \(2n\) group actions (compute \(\tilde{\mathbf Z}_0,\tilde{\mathbf Z}_1\)) + hashing          & \(O(n^{2})\) \\
      \hline
      \(\mathsf{IBBS.Verify}\)  & \(2n\) group actions (recompute \(\tilde{\mathbf Z}_0,\tilde{\mathbf Z}_1\)) + hashing        & \(O(n^{2})\) \\
      \hline
    \end{tabular}
  }\caption{Computational Costs and Asymptotic Complexities}  \label{tab:costs}
\end{table}

In our IBBS scheme, the sizes of all keys and protocol messages grow linearly in the security parameter \(n\).  Concretely, the master public key—which comprises the prime modulus \(p\), the group order \(N\), the base curve \(E_0\), the class‐group generator \(\g\), the hash function description \(\hash\), the \((\text{super})\)–exceptional set \(\mathbf c\in(\Z_N)^n\), and the two master public curves \(\mathbf E_b\in\Ell^n\)—requires \( 2n\bigl\lceil\log_2 p\bigr\rceil \) bits.  The master secret key, consisting of two exponents \(s_0,s_1\in\Z_N\), occupies  $2\,\bigl\lceil\log_2 N\bigr\rceil$ bits.  A user’s secret key \(\mathsf{usk}_{\id}=(\delta,\mathbf x_\delta)\) contains one bit for \(\delta\) plus \(n\) field‐order exponents, for a total of $1+ n\,\bigl\lceil\log_2 N\bigr\rceil$ bits.  The corresponding public key \(\mathsf{pk}_{\id}=(\mathbf X_0,\mathbf X_1)\) is two vectors of \(n\) CSIDH curves, requiring $2n\bigl\lceil\log_2 p\bigr\rceil$ bits.  Finally, the blind signature \(\sigma=(\mathbf c^*_0,\mathbf c^*_1,\mathbf r^*_0,\mathbf r^*_1)\) consists of two ternary vectors of length \(n\) (encoded in \(2n\) bits each) and two integer vectors in \(\Z_N^n\), for a combined size of  $2n\bigl( \lceil\log_2 3\rceil + \lceil\log_2 N\rceil\bigr)=2n\bigl( 2 + \lceil\log_2 N\rceil\bigr)$ bits.  A user identity string is likewise \(n\) bits long.  Thus, every component of the scheme scales only linearly with \(n\), ensuring compact keys and signatures even for high security levels.  Table~\ref{tab:sizes} shows the sizes (in bits) of various components of the IBBS scheme. Table~\ref{tab:ibbs_sizes} represents the sizes of various components of the proposed IBBS scheme at various security levels. 


\begin{table}[h!]
\centering
\resizebox{0.6\textwidth}{!}{
\begin{tabular}{|l|c|}
\hline
\textbf{Components} & \textbf{Sizes in bits} \\
\hline
Master Public Key (MPK)  & $2n\lceil \log_{2} p\rceil$ \\
\hline
Master Secret Key (MSK)  & $2 \lceil \log_{2} N\rceil$ \\
\hline
User Secret Key (USK)  &  $1+ n \lceil \log_{2} N \rceil$ \\
\hline
User Public Key (UPK)  &  $2n\lceil \log_{2} p\rceil$ \\
\hline
Signature ($\sigma$) & $4n+2n\lceil \log_{2} N \rceil ]$ \\
\hline
User Identity ($\mathsf{id}$) & $n$ \\
\hline
\end{tabular}
}\caption{Sizes of various components of the proposed $\mathsf{IBBS}$ scheme}
 \label{tab:sizes}
\end{table}

\begin{table*}[htbp]
  \centering
  \resizebox{\textwidth}{!}{
  \begin{tabular}{|c| c| c| c| c| c| c| c|}
    \hline
    Security Level (bits)
      & $p$ & $n$
      & MPK & MSK
      & USK & UPK
      & Signature \\
    \hline
    80   &  320  &  46  & 29\,440   &   640    & 14\,721  & 29\,440   & 29\,624  \\
    \hline
    100  &  400  &  58  & 46\,400   &   800    & 23\,201  & 46\,400   & 46\,632  \\
    \hline
    128  &  512  &  74  & 75\,776   & 1\,024   & 37\,889  & 75\,776   & 76\,072  \\
    \hline
    192  &  768  & 111  & 170\,496  & 1\,536   & 85\,249  & 170\,496  & 170\,940 \\
    \hline
    256  & 1024  & 148  & 303\,104  & 2\,048   &151\,553  & 303\,104  & 303\,696 \\
    \hline
  \end{tabular}
  } \caption{Bit‐sizes of $\mathsf{IBBS}$ components at various security levels}
    \label{tab:ibbs_sizes}
\end{table*}
Our performance analysis demonstrates that our proposed identity-based blind signature scheme, leveraging the CSIDH-structure, achieves a practical signature size of $9$ KB and $ 37$ KB for 128-bit and 256-bit security, respectively. The computational cost is also reasonable, having complexity of $O(n^2)$ for the various phases of the signature algorithm like Setup, extract, verification, etc.


\section{Conclusion}\label{conclusion}

We have designed an identity-based blind signature protocol that uses the commutative action of the CSIDH group to achieve post-quantum security, and we integrate a zero-knowledge honest verifier sub-protocol to protect the anonymity of the signer and enforce the integrity of the verifier with minimal overhead. Instead of relying on a conventional certificate authority, each user’s public key is directly retrieved from a unique identity string, such as an email address, thus removing the complexity and storage demands of a traditional PKI.  We incorporate blindness through an isogeny-based sigma protocol inspired by Schnorr signatures, which uses constant-size randomness and responses that grow only linearly with the security parameter, yielding compact signatures and linear-scale key material. The scheme’s security is established via a game‐based reduction in the standard model, grounded on the presumed hardness of the GAIP and its multi‐target variant, whose intractability under supersingular isogeny assumptions underpins our proofs. We demonstrate existential unforgeability under adaptive chosen‐message attacks by showing that no efficient antagonist can generate a valid signature on a new message with non‐negligible probability. Additionally, perfect blindness is guaranteed by constructing a simulator for signature transcripts that operates without knowledge of the signer’s secret key, thus preventing any linkage between signatures and signing sessions. Performance measurements confirm that each group action requires only $O(n)$ field operations, so signing, verification, and key extraction all scale linearly with the security parameter. All key and signature components—master keys, user keys, public keys, and signatures—occupy 
$O(n \log p)$  and $O(n \log N)$ bits, which remains practical even for high-security settings. These compact structures and efficient operations make the scheme an excellent fit for resource-constrained environments such as IoT sensors and mobile devices.

Future work includes extending our framework to support dynamic revocation and threshold signing and optimizing CSIDH implementations for side‐channel resistance.

\section*{Conflict of Interest}
The authors declare that they have no conflict of interest.


\begin{thebibliography}{9}

\bibitem{Katsumata2024group}
S. Katsumata, Y.-F. Lai, J. T. LeGrow, and L. Qin, ``CSI-Otter: Isogeny-based (partially) blind signature from the class group action with a twist,'' pp. 729--761, Springer, 2023.

\bibitem{shamir1985identity}
A. Shamir, ``Identity-based cryptosystems and signature schemes,'' in \textit{Advances in Cryptology: Proceedings of CRYPTO 84 4}, pp. 47--53, Springer, 1985.

\bibitem{fiat1986prove}
A. Fiat and A. Shamir, ``How to prove yourself: Practical solutions to identification and signature problems,''  in \textit{Conference on the theory and application of cryptographic techniques}, pp. 186--194, Springer, 1986.

\bibitem{girault1991identity}
M. Girault, ``An identity-based identification scheme based on discrete logarithms modulo a composite number,'' in \textit{Advances in Cryptology--EUROCRYPT'90: Workshop on the Theory and Application of Cryptographic Techniques Aarhus, Denmark, May 21-24, 1990 Proceedings 9}, 
pp. 481-486, Springer, 1991.

\bibitem{tian2013efficient}
M. Tian, L. Huang, and W. Yang, ``Efficient hierarchical identity-based signatures from lattices,'' \textit{International Journal of Electronic Security and Digital Forensics}, 
vol. 5, no. 1, pp. 1--10, 2013.

\bibitem{tian2014efficient}
M. Tian and L. Huang, ``Efficient identity-based signature from lattices,'' 
in \textit{ICT Systems Security and Privacy Protection: 29th IFIP TC 11 International Conference, SEC 2014, Marrakech, Morocco, June 2-4, 2014. Proceedings 29}, 
pp. 321--329, Springer, 2014.

\bibitem{galbraith2017identification}
S. D. Galbraith, C. Petit, and J. Silva, ``Identification protocols and signature schemes based on supersingular isogeny problems,''  in \textit{Advances in Cryptology-ASIACRYPT 2017: 23rd International Conference on the Theory and Applications of Cryptology and Information Security, Hong Kong, China, December 3-7, 2017, Proceedings, Part I 23}, 
pp. 3--33, Springer, 2017.

\bibitem{castryck2018csidh}
W. Castryck, T. Lange, C. Martindale, L. Panny, and J. Renes, ``CSIDH: an efficient post-quantum commutative group action,'' in \textit{Advances in Cryptology-ASIACRYPT 2018: 24th International Conference on the Theory and Application of Cryptology and Information Security, Brisbane, QLD, Australia, December 2-6, 2018, Proceedings, Part III 24}, 
pp. 395--427, Springer, 2018.

\bibitem{de2019seasign}
L. De Feo and S. D. Galbraith, ``SeaSign: compact isogeny signatures from class group actions,'' in \textit{Advances in Cryptology-EUROCRYPT 2019: 38th Annual International Conference on the Theory and Applications of Cryptographic Techniques, Darmstadt, Germany, May 19-23, 2019, Proceedings, Part III 38}, 
pp. 759--789, Springer, 2019.

\bibitem{beullens2019csi}
W. Beullens, T. Kleinjung, and F. Vercauteren, ``CSI-FiSh: efficient isogeny-based signatures through class group computations,'' in \textit{International conference on the theory and application of cryptology and information security}, 
pp. 227--247, Springer, 2019.

\bibitem{peng2020csiibs}
C. Peng, J. Chen, L. Zhou, K.-K. R. Choo, and D. He, ``CSIBS: a post-quantum identity-based signature scheme based on isogenies,'' 
\textit{Journal of Information Security and Applications}, 
vol. 54, p. 102504, 2020.

\bibitem{shaw2021identification}
S. Shaw and R. Dutta, ``Identification scheme and forward-secure signature in identity-based setting from isogenies,'' 
in \textit{International Conference on Provable Security}, 
pp. 309--326, Springer, 2021.

\bibitem{Waterhouse1969}
W. Waterhouse, ``Abelian varieties over finite fields,'' \textit{Annales scientifiques de l'École Normale Supérieure},  vol. 2, pp. 521--560, 1969.

\bibitem{katsumata2024csi}
S. Katsumata, Y.-F. Lai, J. T. LeGrow, and L. Qin, ``CSI-Otter: Isogeny-based (partially) blind signatures from the class group action with a twist,'' \textit{Designs, Codes and Cryptography}, vol. 92, no. 11, pp. 3587--3643, 2024.

\bibitem{atapoor2023csi}
S. Atapoor, K. Baghery, D. Cozzo, and R. Pedersen, 
``CSI-Shark: CSI-FiSh with sharing-friendly keys,'' 
in \textit{Australasian Conference on Information Security and Privacy}, 
pp. 471-502, Springer, 2023.

\bibitem{baghery2021isogeny}
K. Baghery, D. Cozzo, and R. Pedersen, ``An isogeny-based ID protocol using structured public keys,''  in \textit{IMA International Conference on Cryptography and Coding}, 
pp. 179-197, Springer, 2021.

\end{thebibliography}
\end{document}